\documentclass{ws-ijmpe}

\begin{document}

\markboth{A.\ Cucchieri, T.\ Mendes, O.\ Oliveira and P.J.\ Silva}{
Comparing pure Yang-Mills $SU(2)$ and $SU(3)$ propagators}

\catchline{}{}{}{}{}

\title{COMPARING PURE YANG-MILLS $SU(2)$ AND $SU(3)$ PROPAGATORS}

\author{\footnotesize ATTILIO CUCCHIERI and TEREZA MENDES}

\address{Instituto de Fisica de S\~ao Carlos, \\
Universidade de S\~ao Paulo, \\
Caixa Postal 369, 13560-970 S\~ao Carlos, SP, Brazil \\
attilio@ifsc.usp.br, mendes@ifsc.usp.br}

\author{ORLANDO OLIVEIRA and PAULO J. SILVA}

\address{Institute of Physics, University of Coimbra, 3004 516 Coimbra, Portugal\\
orlando@teor.fis.uc.pt, psilva@teor.fis.uc.pt}

\maketitle

\begin{history}
\received{(received date)}
\revised{(revised date)}
\end{history}

\begin{abstract}
The infrared behavior of gluon and ghost propagators in Yang-Mills 
gauge theories is of central importance for the understanding of confinement
in QCD. While analytic studies using Schwinger-Dyson equations predict the
same infrared exponents for the $SU(2)$ and $SU(3)$ gauge groups, 
lattice simulations usually assume that the two cases are different,
although their qualitative infrared features may be the same.
We carry out a comparative study of lattice (Landau) propagators
for both gauge groups. Our data were especially produced with equivalent
lattice parameters to allow a careful comparison of the two cases.
\end{abstract}

\section{\label{intro}Introduction and Motivation}

The study of the infrared limit of QCD is of central importance for the 
comprehension of the mechanisms of quark and gluon confinement and of 
chiral-symmetry breaking. In Landau gauge, the 
Gribov-Zwanziger\cite{GribovZwanziger} and the
Kugo-Ojima\cite{KugoOjima} confinement scenarios predict,
at small momenta, an enhanced ghost propagator and a suppression of the gluon
propagator. The strong infrared divergence for the ghost propagator 
corresponds to a long-range interaction in real space, which may be related to
quark confinement. The suppression of the gluon propagator, which should 
vanish at zero momentum, implies (maximal) violation of reflection positivity 
which may be viewed as an indication of gluon confinement. Analytic studies of
gluon and ghost propagators using Schwinger-Dyson equations (SDE) seem to 
agree with the above scenarios.\cite{SDE} In particular, they predict
for the gluon and for the ghost propagators an infrared exponent
that is independent of the gauge group $SU(N_c)$. 

The Landau-gauge gluon propagator $D(k^2)$ and ghost propagator $G(k^2)$
have been studied by several groups in quenched QCD [i.e.\ pure $SU(3)$ 
Yang-Mills theory], in pure $SU(2)$ Yang-Mills theory (in 2, 3 and 4 
space-time dimensions) and in full
QCD.\cite{Cucchieri,Leinweber,BecirevicBoucaud,MMPreussker,Furui,SilvaOliveira,OliveiraSilva,Maas,violation,Bloch}
In all cases the ghost propagator is enhanced when compared to the tree-level 
behavior $1/k^2$. On the other hand, lattice studies suggest a finite 
nonzero infrared gluon propagator, in contradiction with the infrared 
Schwinger-Dyson solution. One should note that finite-size effects are large 
in the gluon case, as is suggested by investigation of SDE on a 
4-torus\cite{Torus}, and difficult to evaluate. Nevertheless, violation of 
reflection positivity is confirmed by several numerical studies in 3d and in 
4d, for the $SU(2)$ and the $SU(3)$ gauge groups.\cite{violation}

When comparing gluon and ghost propagators from SDE studies to numerical 
results, the agreement is usually at the qualitative level. Moreover, while 
analytic studies using Schwinger-Dyson equations predict the same infrared 
exponents for the $SU(2)$ and $SU(3)$ gauge groups, lattice simulations 
usually assume that the two cases are different, although their qualitative 
infrared features may be the same. In this paper, we carry out a comparative 
study of lattice Landau gauge propagators for these two gauge groups. Our data
were especially produced by considering equivalent lattice parameters in order
to allow a careful comparison of the two cases.

\section{Numerical Simulations and Results}

We consider four different sets of lattice parameters, with the same lattice
size $N^4$ and the same physical lattice spacing $a$ for the two gauge groups.
In particular, in the $SU(3)$ case we used $N^4 = 16^4$ at $\beta = 6.0$,
$N^4 = 24^4$ at $\beta = 6.2$, $N^4 = 32^4$ at $\beta = 6.4$ and
$N^4 = 32^4$ at $\beta = 6.0$. For these cases the lattice spacing
was taken from Ref.\ \refcite{Bali}. The corresponding $\beta$ values
for $SU(2)$ were computed using the asymptotic scaling analysis discussed
in Ref.\ \refcite{Bloch}, yielding $\beta = 2.4469, 2.5501$ and $2.6408$.
In the first three cases the physical lattice volume $V = (Na)^4$ is 
approximately the same, i.e.\ $V \approx (1.7$ fm$)^4$. The fourth case 
corresponds to a significantly larger physical volume, i.e.\ 
$V \approx (3.2$ fm$)^4$.

The gluon propagator $D(k)$ and the ghost propagator $G(k)$ were evaluated
as a function of the lattice momentum $k$, using 50 configurations for all
four cases. In order to compare the propagators from the different 
simulations, the propagators were renormalized to their tree-level value 
$1 / \mu^2$, using $\mu = 3$ GeV as a renormalization point.
Details of the simulations can be found in Ref.\ \refcite{Cucchieri:2007zm}.

\begin{figure}[th]
\centerline{\psfig{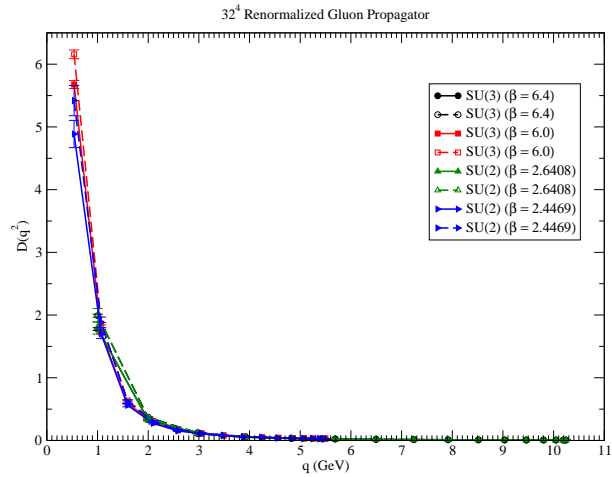}}
\vspace*{8pt}
\caption{Renormalized gluon propagator for $32^4$ lattices. The solid line is
the propagator computed using the renormalization constant $Z$ associated with
momenta $(k,0,0,0)$. The dashed line uses $Z$ associated with $(k,k,0,0)$
momenta. In this figure, the data report only momenta of $(k,k,0,0)$ 
type. \label{finiteffects}}
\end{figure}

In what concerns finite-volume and finite-lattice-spacing effects,
we find a slight dependence on the type of momenta 
[$(k,0,0,0)$, $(k,k,0,0)$, $(k,k,k,0)$ or $(k,k,k,k)$]
where the renormalization is performed. Fig.\ \ref{finiteffects} 
illustrates the effects of choosing different renormalization constants for
the gluon propagator. From now on, we will consider only the data computed
using renormalization constants for momenta $(k,0,0,0)$.

\begin{figure}[th]
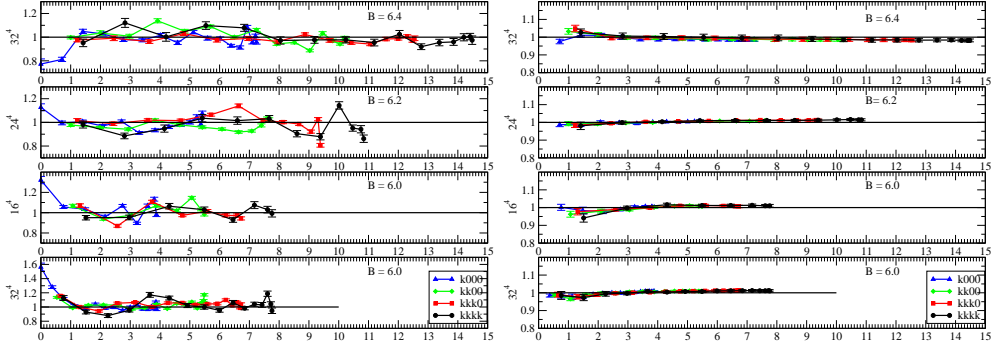

\centerline{\psfig{file=ratios.glue2.eps,width=6.5cm}
            \psfig{file=ratios.ghost2.eps,width=6.5cm}}
\vspace*{8pt}
\caption{Ratios of $SU(3)$ over $SU(2)$ gluon (left) and ghost (right)
propagators (for the four lattice setups) considered as a function of the 
magnitude $k$ of the four momentum in GeV. \label{ratiosglue}}
\end{figure}

In Fig.\ \ref{ratiosglue} we show the ratios of the renormalized $SU(3)$ over 
$SU(2)$ propagators for various lattice setups. The data show that the two 
cases have very similar finite-size and discretization effects. Moreover, 
$SU(2)$ and $SU(3)$ propagators are essentially equal, with slight differences
in the low-momenta region (especially for the gluon propagator). Thus, our 
study supports the prediction from the Schwinger-Dyson equations that the 
propagators are the same for all $SU(N_c)$ groups in the nonperturbative 
region. Clearly, further studies are required before drawing conclusions about
the comparison between $SU(2)$ and $SU(3)$ propagators in the deep-infrared 
region, where the gluon propagator may show a turnover and a suppression, as 
predicted in the Gribov-Zwanziger scenario.

\section*{Acknowledgements}

The authors thank R. Alkofer, A. Maas and C. Fischer for discussions.
O.O.\ and P.J.S.\ acknowledge FCT for financial support under contract
POCI/FP/63923/2005. P.J.S. acknowledges financial support from FCT via grant
SFRH/BD/10740/2002. O.O. was also supported by FAPESP (grant \# 06/61514-8) 
during his stay at IFSC-USP.
A.C.\ and T.M.\ were supported by FAPESP and by CNPq.
Parts of our simulations have been done on the IBM supercomputer
at S\~ao Paulo University (FAPESP grant \# 04/08928-3) and on the
supercomputer Milipeia at Coimbra University.

\end{document}